\documentclass[fleqn,twoside]{article}

\usepackage{mathptmx}
\usepackage{epsfig}
\usepackage{booktabs}
\usepackage{dcolumn}
\usepackage{amsmath}
\usepackage{psfrag}
\usepackage{espcrc2}

\newcommand{\hb}{\mbox{HERA-B}}

\newcommand{\adepp}{$A$-dependence}
\newcommand{\adep}{\adepp{} }
\newcommand{\pA}{\ensuremath{\mathrm{p}A}}
\newcommand{\pN}{\ensuremath{\mathrm{p}N}}

\newcommand{\de}{\ensuremath{\mathrm{d}}}

\newcommand{\unit}[1]{\ensuremath{\:\mathrm{#1}}}

\newcommand{\gevc}{\ensuremath{\unit{GeV}/c}}

\newcommand{\xf} {{\ensuremath{x_\mathrm{F}}}}
\newcommand{\pt} {{\ensuremath{p_\mathrm{T}}}}

\newcommand{\cquark}{{\ensuremath{\mathrm{c}}}}

\newcommand{\PJgy}{{\ensuremath{\mathrm{J}\mskip -2mu/\mskip -2mu\psi}}}
\newcommand{\jpsi}{\PJgy}
\newcommand{\jpsibf}{\ensuremath{\mathbf{J \mskip -2mu/\mskip -2mu\psi}}}
\newcommand{\psiprime}{{\ensuremath{\psi}'}}
\newcommand{\psitwos}{{\ensuremath{\psi(2S)}}}
\newcommand{\chic}{{\ensuremath{\chi_\cquark}}}
\newcommand{\ccbar}{{\ensuremath{\mathrm{c\overline{c}}}}}

\newcommand{\PKp} {{\ensuremath{\mathrm{K}^+}}}                    
\newcommand{\PKm} {{\ensuremath{\mathrm{K}^-}}}                    

\newcommand{\PDp}{{\ensuremath{\mathrm{D^+}}}}                     
\newcommand{\PDstarp}{{\ensuremath{\mathrm{D^{*+}}}}}              
\newcommand{\PDz} {{\ensuremath{\mathrm{D^0}}}}                    

\newcommand{\epem}{{\ensuremath{\mathrm{e^+ e^-}}}}                         
\newcommand{\dielectron}{\epem}
\newcommand{\mpmm}{\ensuremath{\mu^+ \mu^-}}                  
\newcommand{\dimuon}{\mpmm}

\newcommand{\jpmm}{{\ensuremath{\jpsi\rightarrow\dimuon}}}

\newcommand{\jpee}{{\ensuremath{\jpsi\rightarrow\dielectron}}}

\DeclareFontFamily{OML}{mygreek}{\skewchar \font =127}
\DeclareFontShape{OML}{mygreek}{m}{rm}{<-> [0.91] zptmcm7m }{}
\DeclareSymbolFont{lettersgreek}{OML}{mygreek}{m}{rm}

\DeclareMathSymbol{\alpha}{0}{lettersgreek}{"0B}
\DeclareMathSymbol{\beta}{0}{lettersgreek}{"0C}
\DeclareMathSymbol{\gamma}{0}{lettersgreek}{"0D}
\DeclareMathSymbol{\delta}{0}{lettersgreek}{"0E}
\DeclareMathSymbol{\epsilon}{0}{lettersgreek}{"0F}
\DeclareMathSymbol{\zeta}{0}{lettersgreek}{"10}
\DeclareMathSymbol{\eta}{0}{lettersgreek}{"11}
\DeclareMathSymbol{\theta}{0}{lettersgreek}{"12}
\DeclareMathSymbol{\iota}{0}{lettersgreek}{"13}
\DeclareMathSymbol{\kappa}{0}{lettersgreek}{"14}
\DeclareMathSymbol{\lambda}{0}{lettersgreek}{"15}
\DeclareMathSymbol{\mu}{0}{lettersgreek}{"16}
\DeclareMathSymbol{\nu}{0}{lettersgreek}{"17}
\DeclareMathSymbol{\xi}{0}{lettersgreek}{"18}
\DeclareMathSymbol{\pi}{0}{lettersgreek}{"19}
\DeclareMathSymbol{\rho}{0}{lettersgreek}{"1A}
\DeclareMathSymbol{\sigma}{0}{lettersgreek}{"1B}
\DeclareMathSymbol{\tau}{0}{lettersgreek}{"1C}
\DeclareMathSymbol{\upsilon}{0}{lettersgreek}{"1D}
\DeclareMathSymbol{\phi}{0}{lettersgreek}{"1E}
\DeclareMathSymbol{\chi}{0}{lettersgreek}{"1F}
\DeclareMathSymbol{\psi}{0}{lettersgreek}{"20}
\DeclareMathSymbol{\omega}{0}{lettersgreek}{"21}
\DeclareMathSymbol{\varepsilon}{0}{lettersgreek}{"22}
\DeclareMathSymbol{\vartheta}{0}{lettersgreek}{"23}
\DeclareMathSymbol{\varomega}{0}{lettersgreek}{"24}
\DeclareMathSymbol{\varrho}{0}{lettersgreek}{"25}
\DeclareMathSymbol{\varsigma}{0}{lettersgreek}{"26}
\DeclareMathSymbol{\varphi}{0}{lettersgreek}{"27}

\title{Open and Hidden Charm Production in 920 GeV Proton-Nucleus Collisions}
\author{U.\ Husemann\address[Siegen]{Fachbereich Physik, Universit{\"a}t Siegen, Walter-Flex-Str.\ 3, D--57068 Siegen, Germany}
 for the \hb{} Collaboration}

\begin{document}

\begin{abstract}
The \hb{} collaboration has studied the production of charmonium and 
open charm states in
collisions of 920\unit{GeV} protons with wire targets of different materials.
The acceptance of the \hb{} spectrometer covers negative values of \xf{} up 
to $\xf=-0.3$ and a broad range in transverse momentum from 
0.0 to 4.8\gevc. 
The studies presented in this paper
include \jpsi{} differential distributions 
and the suppression of \jpsi{} production in nuclear media. Furthermore,
production cross sections and cross section
ratios for open charm mesons are discussed.

\vspace{-60mm}

\begin{flushright}
\begin{minipage}{0.3\textwidth}
 \flushright\normalsize SI-HEP-2004-10
\end{minipage}
\end{flushright}
\vspace{47mm}

\end{abstract}

\maketitle

\section{Introduction}
Studying charmonium and open charm hadroproduction provides a good test of
the theoretical production models available and their modifications in
nuclear media. Since many of these models rely on experimental data to 
adjust their free parameters, a comprehensive test of the model
predictions in a broad kinematic range and with good precision is desirable.

HERA-B is in the position to extend previous measurements of \jpsi{}
production and nuclear suppression to negative \xf{} values.
\PDz{} and \PDp{} production cross sections and cross section ratios
are determined with high accuracy.

\section{The \hb{} Detector and Trigger}
\begin{figure}[ht]
\centering
\includegraphics[width=0.47\textwidth]{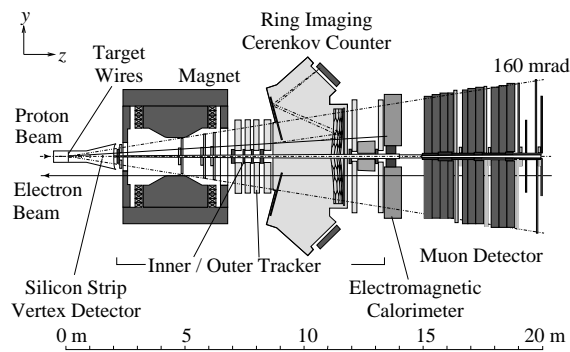}
\vspace{-10mm}

\caption{Side view of the \hb{} detector.}
\label{fig:detector}
\end{figure}

The \hb{} detector is a fixed-target spectrometer with large angular 
acceptance. The detector includes
subdetectors for vertexing, tracking, and particle identification, as shown in
Fig.~\ref{fig:detector}.

Protons with energies of 920\unit{GeV} are brought into collisions
with an internal wire target. The target consists of two stations of four
wires each, separated by 4\unit{cm} along the beam axis.
Each wire can be moved independently in the halo of the proton beam to adjust
the interaction rate.

The vertex detector system (VDS) allows for reconstruction and 
separation of primary and  secondary vertices. It consists of eight 
double-layers of double-sided
silicon strip detectors. The tracking detector is divided into an
inner part consisting of micro-strip gaseous chambers with
gas electron multiplier foils and an outer part of honeycomb
drift chambers.

The particle identification devices include a ring-imaging \v Cerenkov 
counter (RICH),
an  electromagnetic calorimeter (ECAL),
and a four-layer muon detector.

In \hb, a multilevel trigger system is employed to enrich lepton pairs
from  \jpsi{} decays. 
Pretriggers in the muon detector and ECAL provide track seeds for the
first level trigger (FLT). The FLT performs a 
track search in the outer tracker. 
In the second trigger level, the tracking of
the FLT is confirmed, tracks are extrapolated to 
the VDS, and a vertex fit is performed for pairs of leptons. Accepted events
are reconstructed online.
Alternatively, a minimum bias trigger can be utilized, 
requiring either a minimum number of 
photons in the RICH or a minimum energy deposit in the ECAL.

\section{The 2002/2003 Data-Taking Period}
The results presented in this paper are based on data taken
with the \hb{} detector from November 2002 to February 2003. 
In dilepton triggered data, about 170,000 \jpmm{} decays and 
about 150,000 \jpee{} decays have been reconstructed. In addition,
about 200 million events have been recorded with the minimum bias trigger.

\section{Charmonium Production}

%
\subsection{\jpsibf{} Cross Section}
\begin{figure}[t]
\centering
\includegraphics[width=0.23\textwidth]{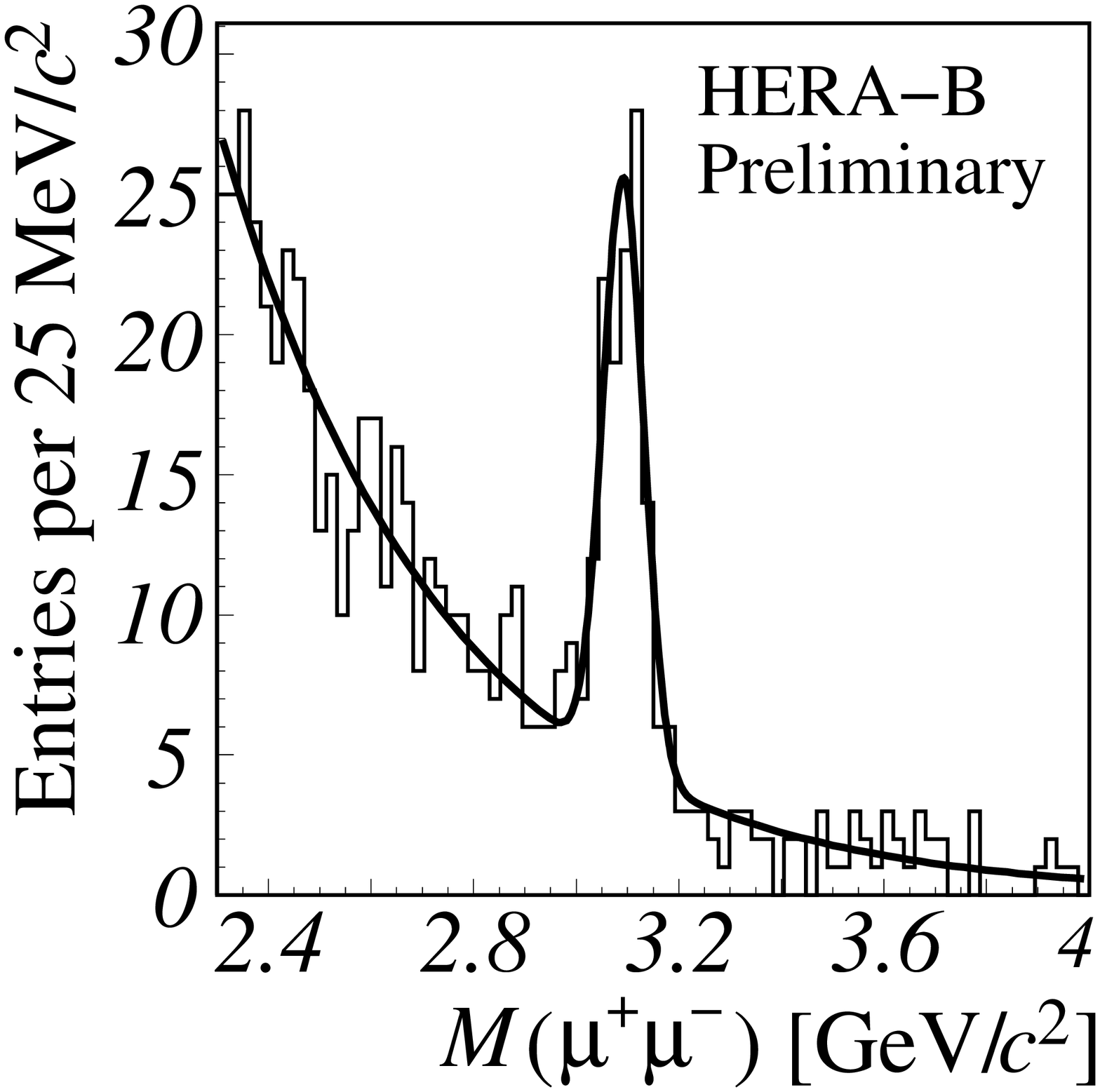}
\includegraphics[width=0.23\textwidth]{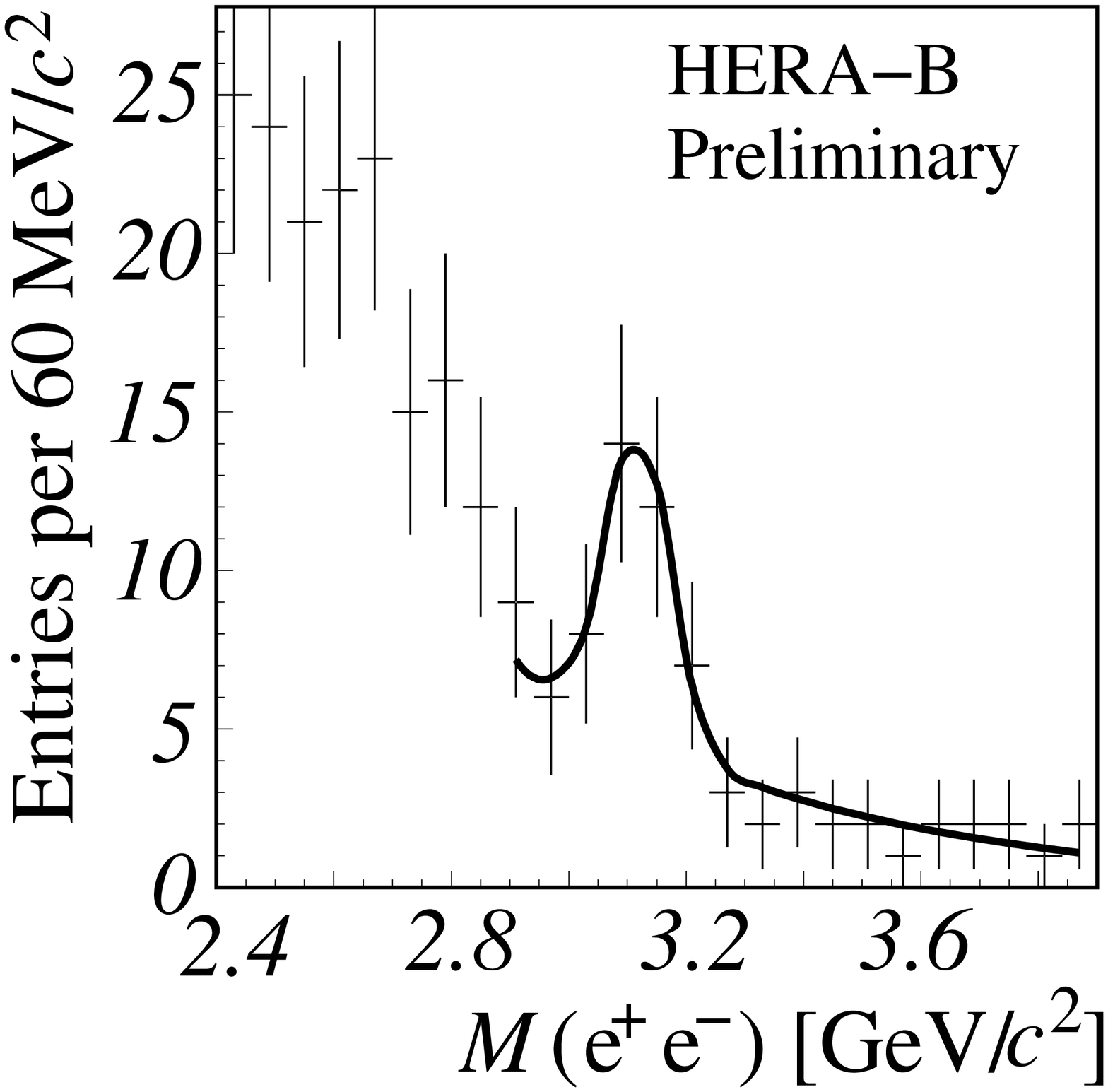}
\vspace{-14mm}

\caption{\jpsi{} signal in \jpmm{} decays (left) and from \jpee{} 
decays (right), extracted from minimum bias triggered data.}
\label{fig:jpsi}
\end{figure}

%
A clean \jpsi{} signal is observed in the data recorded with the minimum
bias trigger. The invariant mass distributions for the dimuon and the
dielectron channels are shown in Fig.~\ref{fig:jpsi}. From these signals 
the \jpsi{} production cross section can be extracted without trigger bias.

%
\subsection{\jpsibf{} Differential Distributions}
\begin{figure}[t]
\centering
\includegraphics[width=0.28\textwidth]{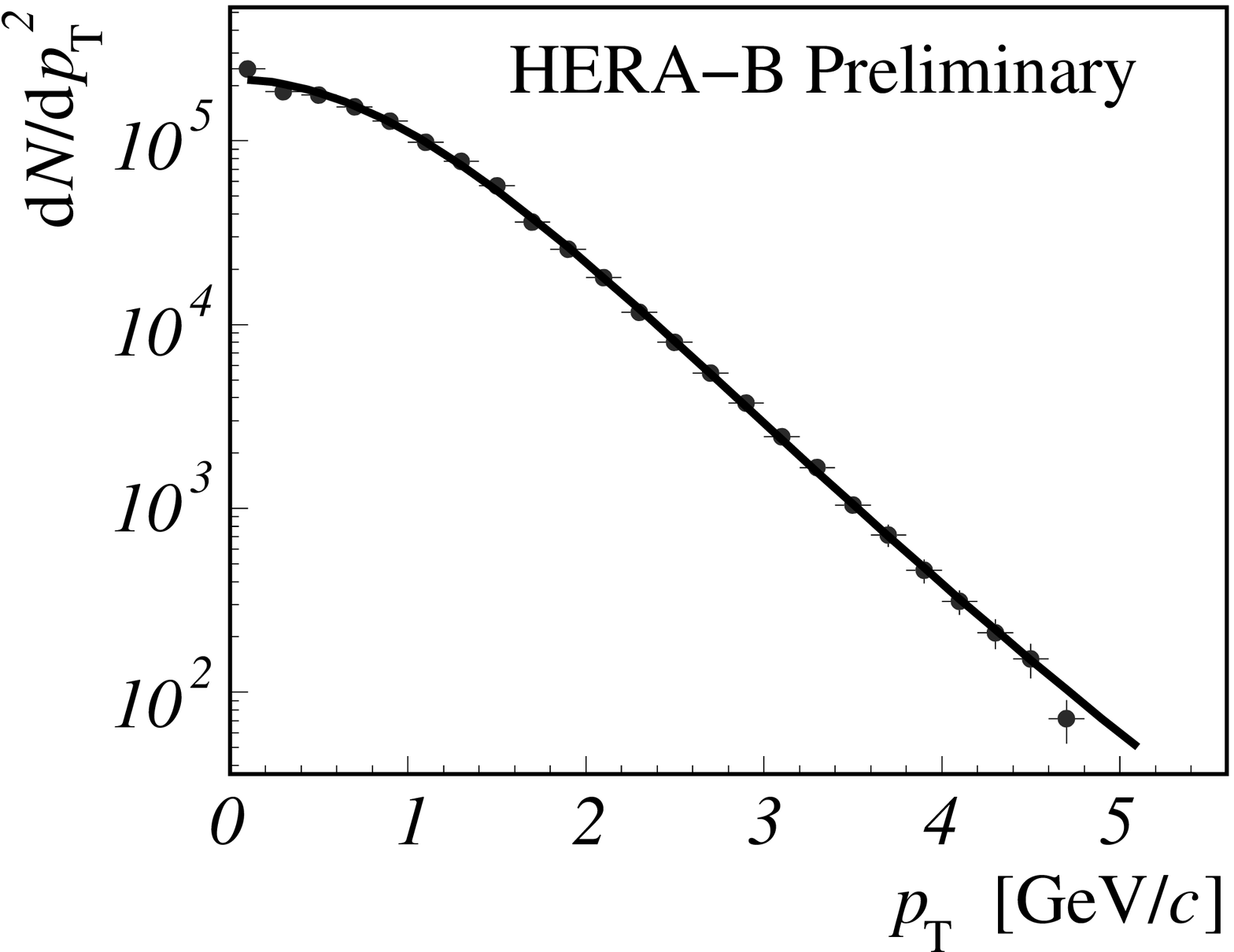}
\vspace{-10mm}

\caption{Transverse momentum  distribution $\de N/\de\pt^2$ of \jpee{} decays.}
\label{fig:pt}
\end{figure}

%
From data recorded with the dilepton trigger, differential
cross sections of \jpsi{} production are determined.
As an example, in Fig.~\ref{fig:pt}, the \jpee{} yield for 
25\% of the full  statistics is shown as a function of the 
transverse momentum \pt{}.
The \hb{} detector covers a broad kinematic range of 0.0 to 4.8\gevc.
The \pt{} spectrum is fitted with the parametrization function
\begin{equation}
\frac{\de\sigma}{\de\pt^2} = A\cdot\left[ 1 + 
  \left( \frac{ 35\pi\, \pt }{256\,\langle \pt \rangle} \right)^2\right]^{-6},
\end{equation}
in which the normalization constant $A$ and the average transverse 
momentum $\langle \pt \rangle$ are free parameters. 
In Table~\ref{table:pt}, preliminary HERA-B results on $\langle \pt \rangle$  
are shown. The comparison of different target materials and with results
of the experiments E771~\cite{Alexopoulos:1997yd} 
and E789~\cite{Schub:1995pu} shows an increase of $\langle \pt \rangle$ 
with the atomic mass number.

\begin{table*}
\caption{HERA-B preliminary results on the average transverse momentum 
$\langle \pt \rangle$ for \jpee{} and \jpmm{} decays, compared to
results of previous experiments.}
\label{table:pt}
\begin{center}
\small
\begin{tabular}{ccccc}
\toprule
Target & Experiment & Max.\ $p_\mathrm{T}$ & $\langle p_\mathrm{T}\rangle [\mathrm{GeV}/c] (\mathrm{e^+ e^-})$ & $\langle p_\mathrm{T}\rangle [\mathrm{GeV}/c]  (\mu^+ \mu^-)$ \\
\midrule
C, 920\,GeV & HERA-B Preliminary& $4.8\:\mathrm{GeV}/c$ & $1.22 \pm 0.01\mathrm{(stat.)}$ & $1.22 \pm 0.01\mathrm{(stat.)}$ \\
W, 920\,GeV & HERA-B Preliminary& $4.8\:\mathrm{GeV}/c$ & $1.29 \pm 0.01\mathrm{(stat.)}$ & $1,30 \pm 0.01\mathrm{(stat.)}$ \\
Si, 800\,GeV & E771~\cite{Alexopoulos:1997yd} & $3.5\:\mathrm{GeV}/c$ & & $1.20 \pm 0.01$\\
Au, 800\,GeV & E789~\cite{Schub:1995pu} & $2.6\:\mathrm{GeV}/c$ & & $1.290 \pm 0.009$ \\
\bottomrule
\end{tabular}
\end{center}
\end{table*}

%
\subsection{$\mathbf{A}$-Dependence of \jpsibf{} Production}
%

Nuclear effects in heavy quark production are commonly parametrized by 
the power law
\begin{equation}
  \sigma_{\pA} = \sigma_{\pN} \cdot A^{\alpha(\xf,\pt)}.
  \label{eq:suppression}
\end{equation}
Here, $\sigma_{\pA}$ is the proton-nucleus cross section, $\sigma_{\pN}$ is the
proton-nucleon cross section, and $A$ is the atomic mass number of the target.
A value of \mbox{$\alpha(\xf,\pt)<1$} corresponds to suppression of
heavy quark production. The sources of nuclear suppression in
\jpsi{} production are either
interactions of the partons in the initial state with the nuclear medium,
or the absorption of the \ccbar{} pre-state or the final \jpsi, see
e.g.~\cite{Vogt:1999dw}.

In \hb, the suppression parameter $\alpha$ is extracted from data samples
taken with a carbon and a tungsten target
simultaneously. This setup is utilized to minimize systematic uncertainties
in the detector and trigger performance. 
Using $\sigma=N/(\mathcal{L}\varepsilon)$, Eq.~(\ref{eq:suppression}) can
be solved for $\alpha$:
\begin{equation}
  \alpha = \frac{1}{\log\left(A_\mathrm{W}/A_\mathrm{C}\right)}
  \cdot \log\left( 
    \frac{N_\mathrm{W}}{N_\mathrm{C}}\cdot
    \frac{\mathcal{L}_\mathrm{C}}{\mathcal{L}_\mathrm{W}}\cdot
    \frac{\varepsilon_\mathrm{C}}{\varepsilon_\mathrm{W}}
  \right).
\end{equation}

\begin{figure}[t]
\centering
\includegraphics[width=0.3\textwidth]{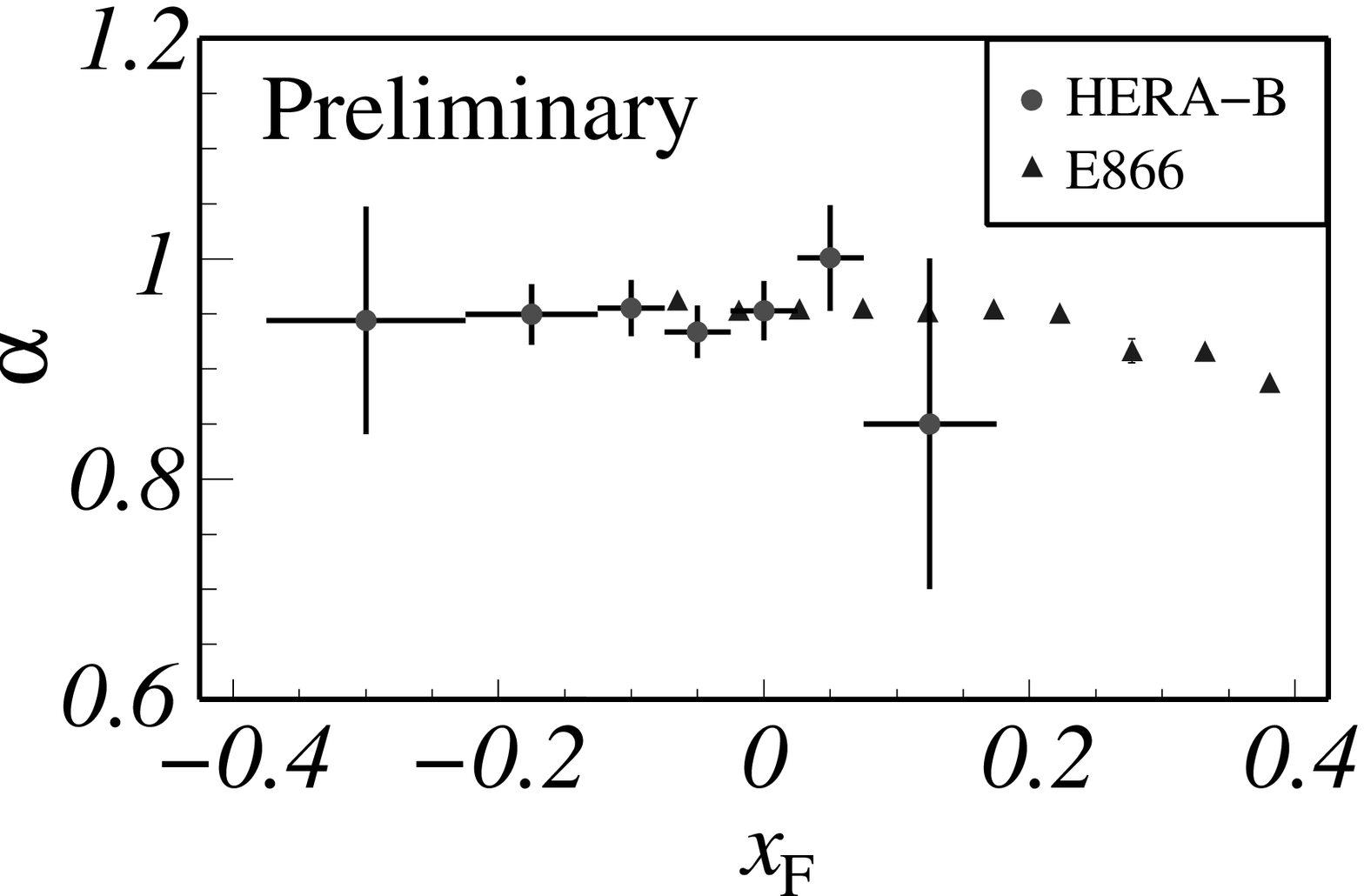}
\\[1mm]
\includegraphics[width=0.3\textwidth]{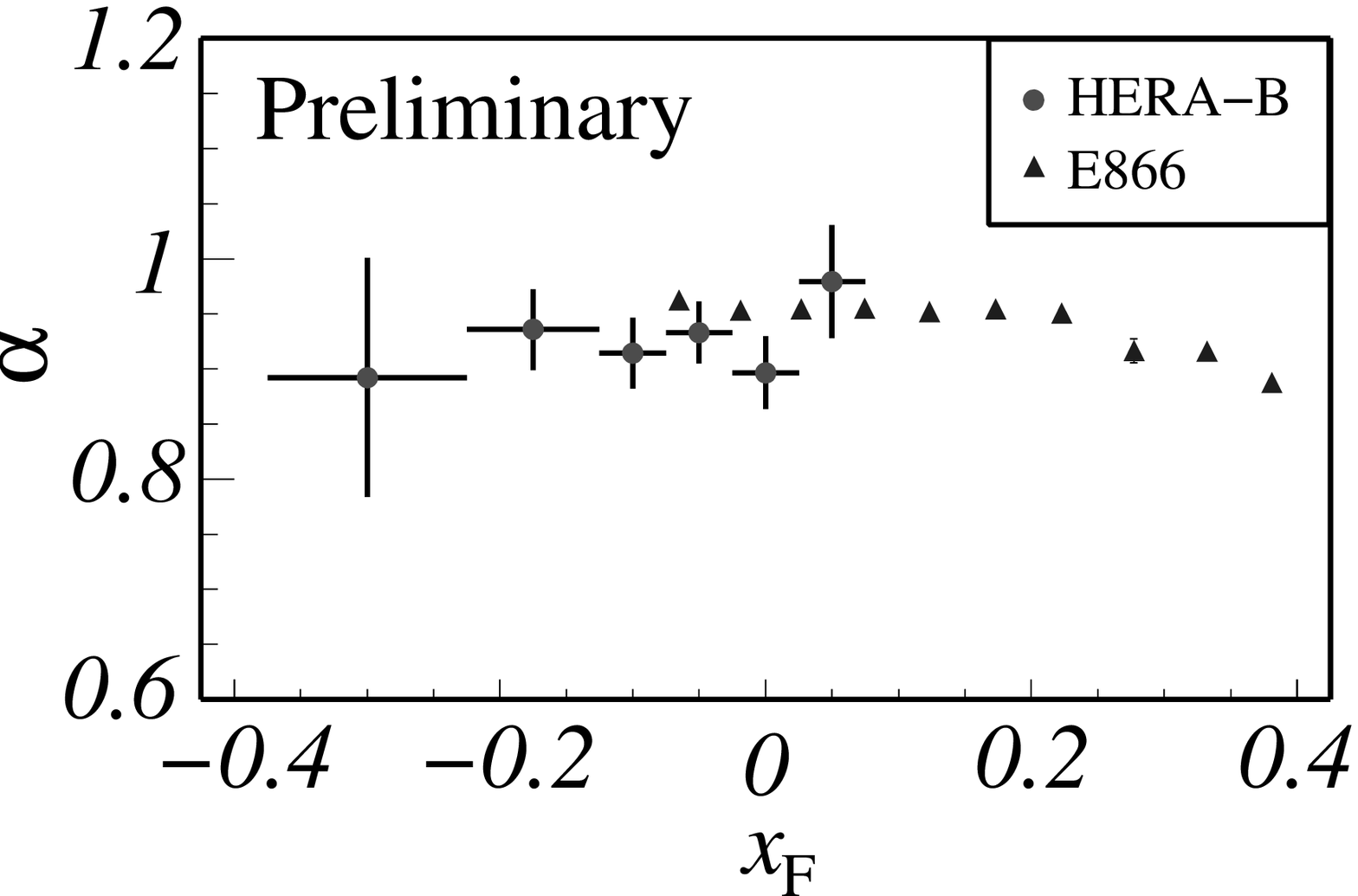}
\vspace{-8mm}

\caption{Nuclear suppression parameter $\alpha(\xf)$ for two different
combinations of carbon and tungsten wires, compared to the result
of E866~\cite{Leitch:1999ea}.}
\label{fig:alpha}
\end{figure}

The \adep measurement hence includes measuring three ratios, the
ratio of \jpsi{} yields, $N_\mathrm{W}/N_\mathrm{C}$, the ratio of 
luminosities per wire, $\mathcal{L}_\mathrm{C}/\mathcal{L}_\mathrm{W}$,
and the ratio of efficiencies,
$\varepsilon_\mathrm{C}/\varepsilon_\mathrm{W}$. 
The luminosity ratio is extracted from the number of interactions
in minimum bias events recorded in parallel to the triggered data.
The ratio of efficiencies is determined from a detailed Monte Carlo (MC) 
simulation of the \hb{} detector and trigger.
Preliminary results on $\alpha(\xf)$, based on 10\% of the full statistics,
are presented in Fig.~\ref{fig:alpha}.
Data from different combinations of carbon and tungsten wires 
show results consistent with the previous result by the E866 
collaboration~\cite{Leitch:1999ea} in the overlap region
and allow the measurement to be extended
to $\xf=-0.3$.

%
\subsection{\psitwos{} Production}
%
The production cross section for \psitwos{} mesons is measured relative
to the \jpsi{} cross section in order to reduce systematic uncertainties.
In order to extract 
\begin{equation}
R=\frac{\mathrm{BR}(\psitwos \rightarrow \ell^+ \ell^-)\cdot\sigma_\psitwos}
  {\mathrm{BR}(\jpsi \rightarrow \ell^+ \ell^-)\cdot\sigma_\jpsi}
  = \frac{N_\psitwos}{N_\jpsi}\cdot
  \frac{\varepsilon_\jpsi}{\varepsilon_\psitwos},
\end{equation}
the relative yield $N_\psitwos/N_\jpsi$ is corrected by the efficiency
ratio $\varepsilon_\jpsi/\varepsilon_\psitwos$, which is obtained
from a MC simulation.

\begin{figure}[t]
\centering
\includegraphics[width=0.37\textwidth]{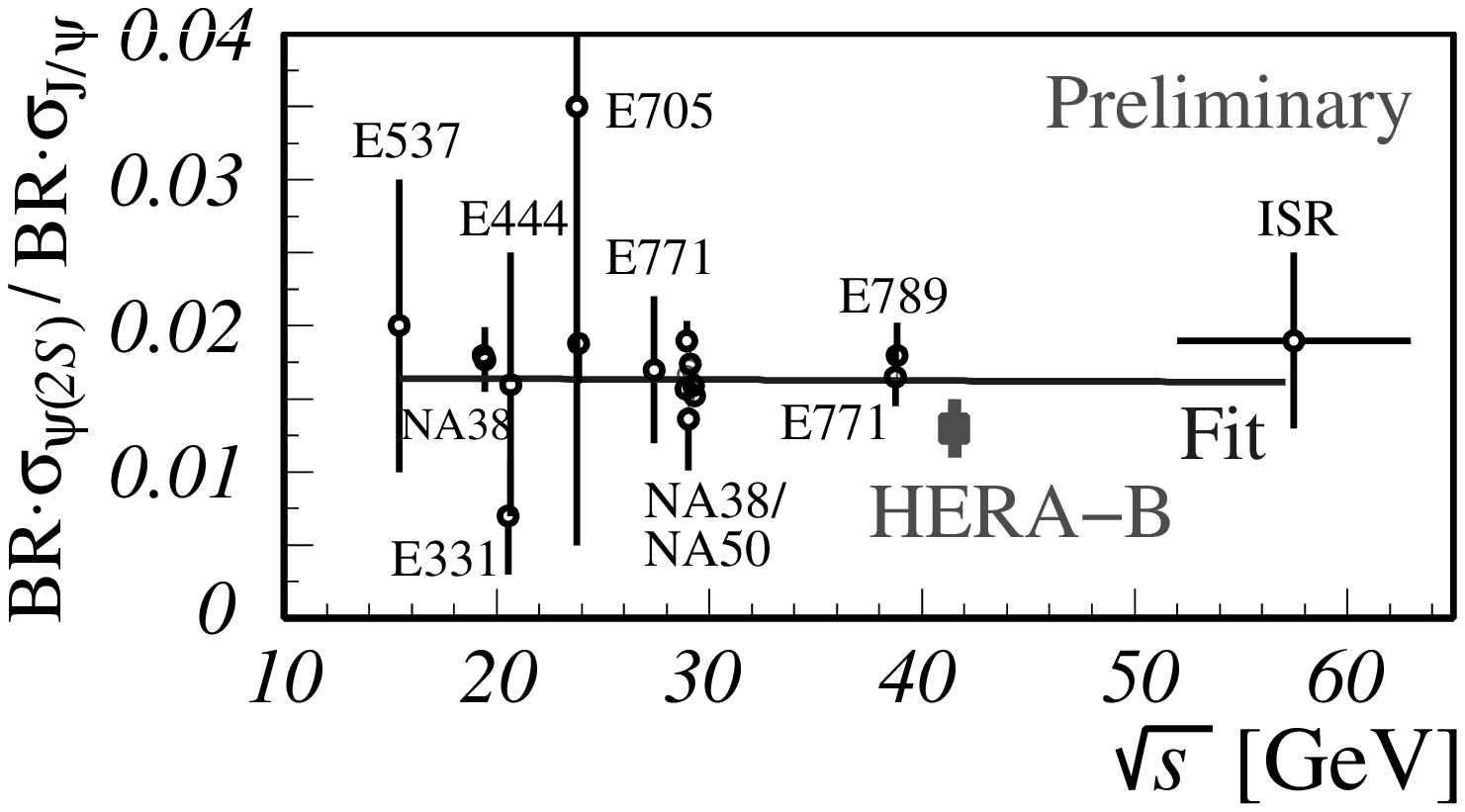}
\includegraphics[width=0.37\textwidth]{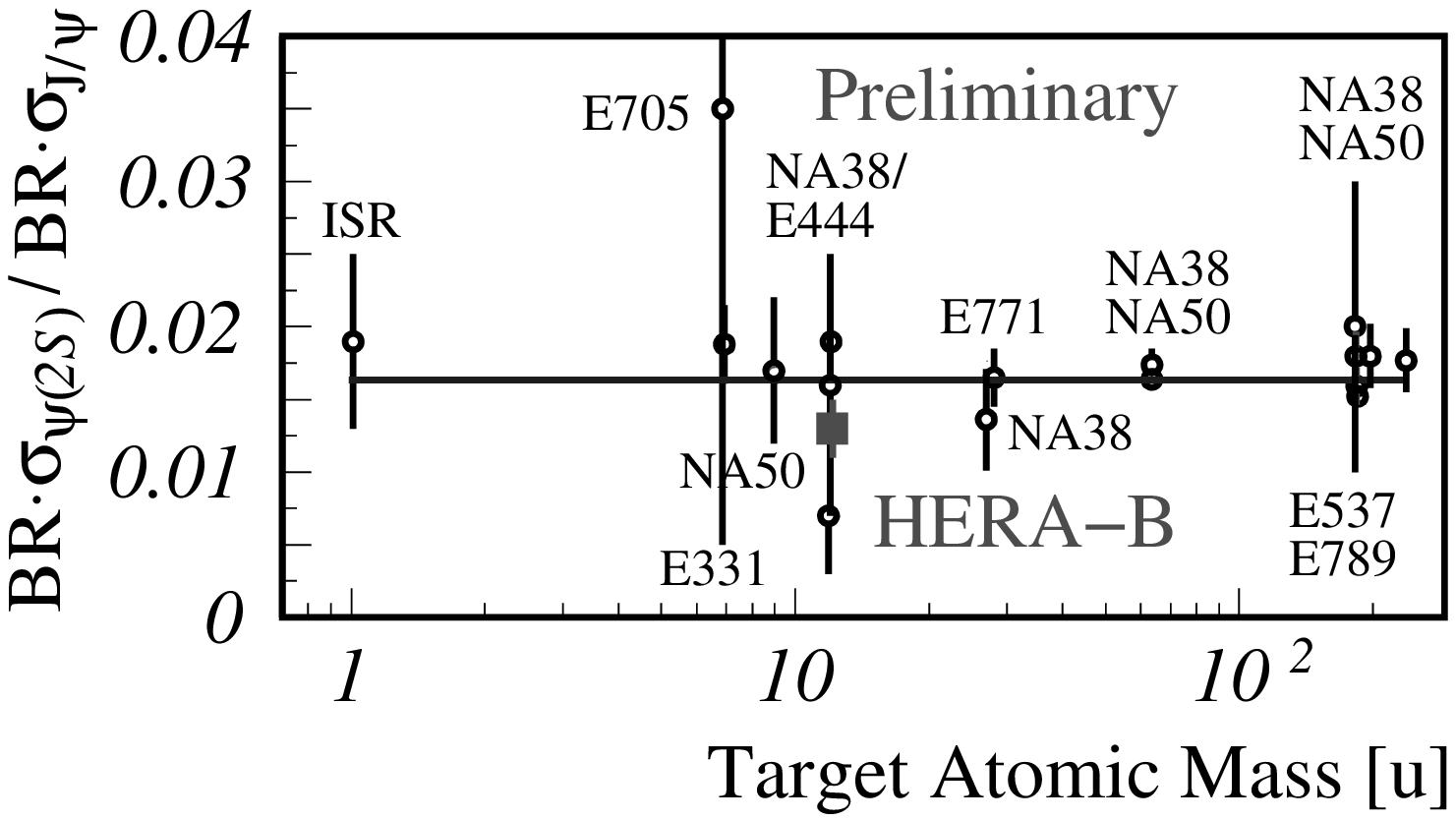}
\vspace{-8mm}

\caption{\psitwos-to-\jpsi{} production ratio as a function of the
  center-of-mass energy  $\sqrt{s}$ (top) and the 
  atomic mass number $A$ (bottom). The preliminary \hb{} result 
(square) is compared to results of previous experiments.}
\label{fig:psipratio}
\end{figure}

A preliminary result for 
the \epem{} channel is $R=0.13\pm0.02(\mathrm{stat.})$.
As shown in Fig.~\ref{fig:psipratio}, this result
agrees well with previous measurements, both as a function of
the center-of-mass energy and the atomic mass number.

%
\subsection{\chic{} Production}
\begin{figure}[t]
\centering
\includegraphics[width=0.30\textwidth]{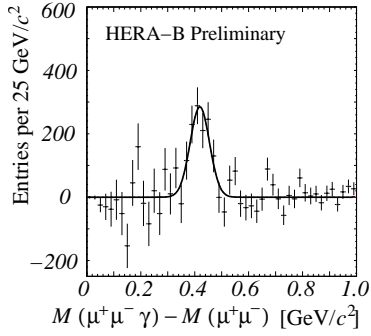}
\vspace{-8mm}

\caption{\chic{} signal in mass difference distribution 
$M(\dimuon\gamma)-M(\dimuon)$. Combinatorial background has been subtracted.}
\label{fig:chic}
\end{figure}

%
The fraction of \jpsi{} produced via the radiative decay 
$\chi_{\mathrm{c}1,2} \rightarrow \jpsi\,\gamma\rightarrow \dimuon\gamma$, 
\begin{equation}
R(\chic) = \frac{\sum_{J=1}^2 \sigma_{\chi_{\cquark J}} \cdot 
  \mathrm{BR}(\chi_{\cquark J} \rightarrow \jpsi\gamma)}{\sigma_\jpsi} =
\frac{N_\chic}{N_\jpsi}\cdot
\frac{\varepsilon_\jpsi}{\varepsilon_\chic\cdot\varepsilon_\gamma},
\end{equation}
provides a good test of charmonium production models. As depicted
in Fig.~\ref{fig:chic}, the \chic{}
signal is extracted from the mass difference distribution of the 
$\dimuon\gamma$ and the $\dimuon$ systems. The efficiency $\varepsilon_\gamma$
to detect the additional photon is determined from a MC simulation.

A preliminary analysis of
a part of the 2002/2003 data yields $R(\chic)=0.21\pm 0.5(\mathrm{stat.})$.
This result is consistent with the previous \hb{} result,
$R(\chic)=0.32\pm 0.6(\mathrm{stat.})\pm 0.4(\mathrm{syst.})$~\cite{Abt:2002vq}.

\section{Open Charm Production}

%
\subsection{D Meson Production Cross Sections}
\begin{figure*}
\centering
{\psfrag{events}{\small $189\pm20$ ev}
\psfrag{xaxis}{\small $M(\PKm\pi^+)$ [GeV/$c^2$]}
\includegraphics[width=0.3\textwidth]{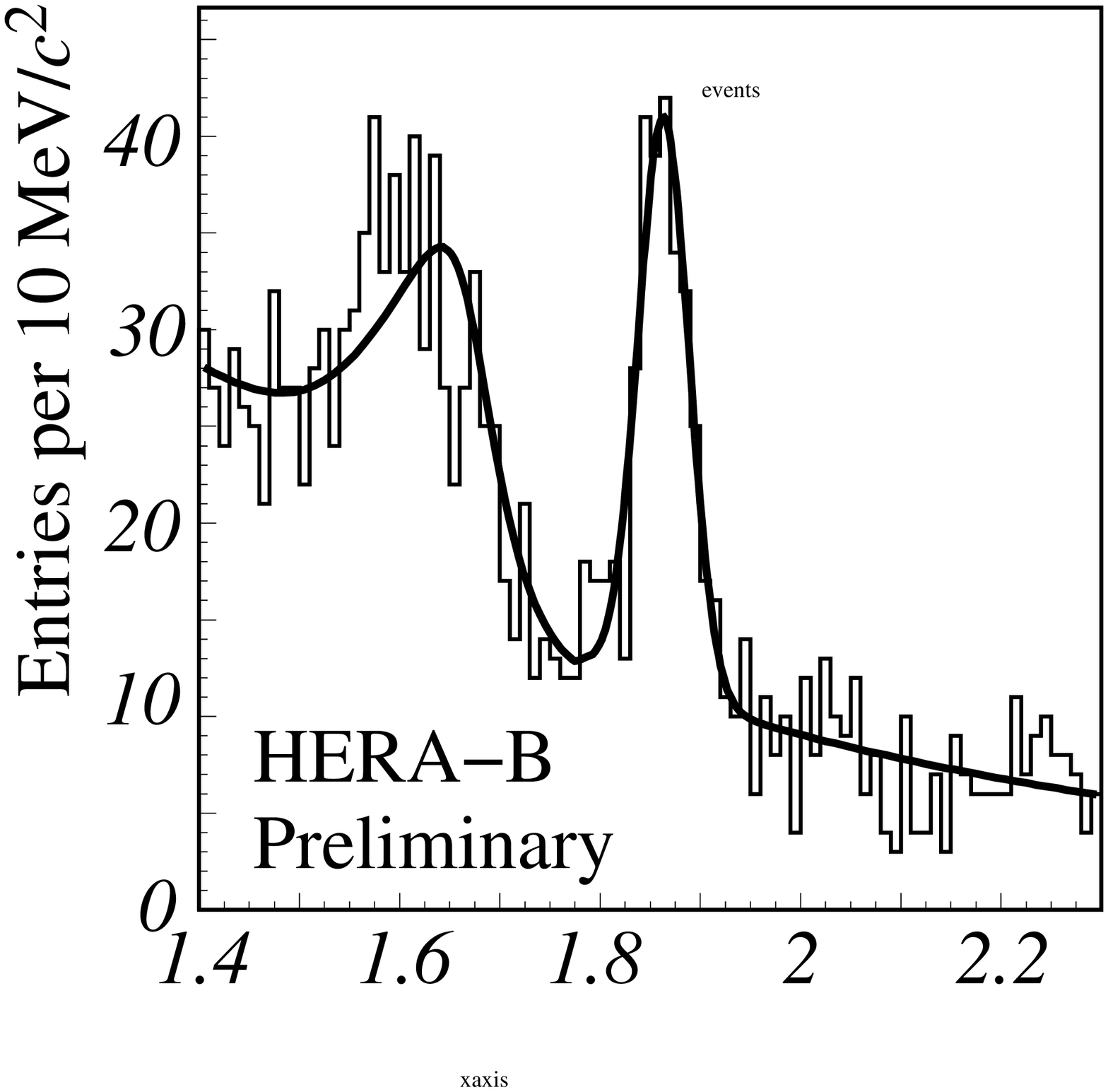}}
\hspace{3mm}
{\psfrag{events}{\small $98\pm12$ ev}
\psfrag{xaxis}{\small $M(\PKm\pi^+\pi^+)$ [GeV/$c^2$]}\includegraphics[width=0.3\textwidth]{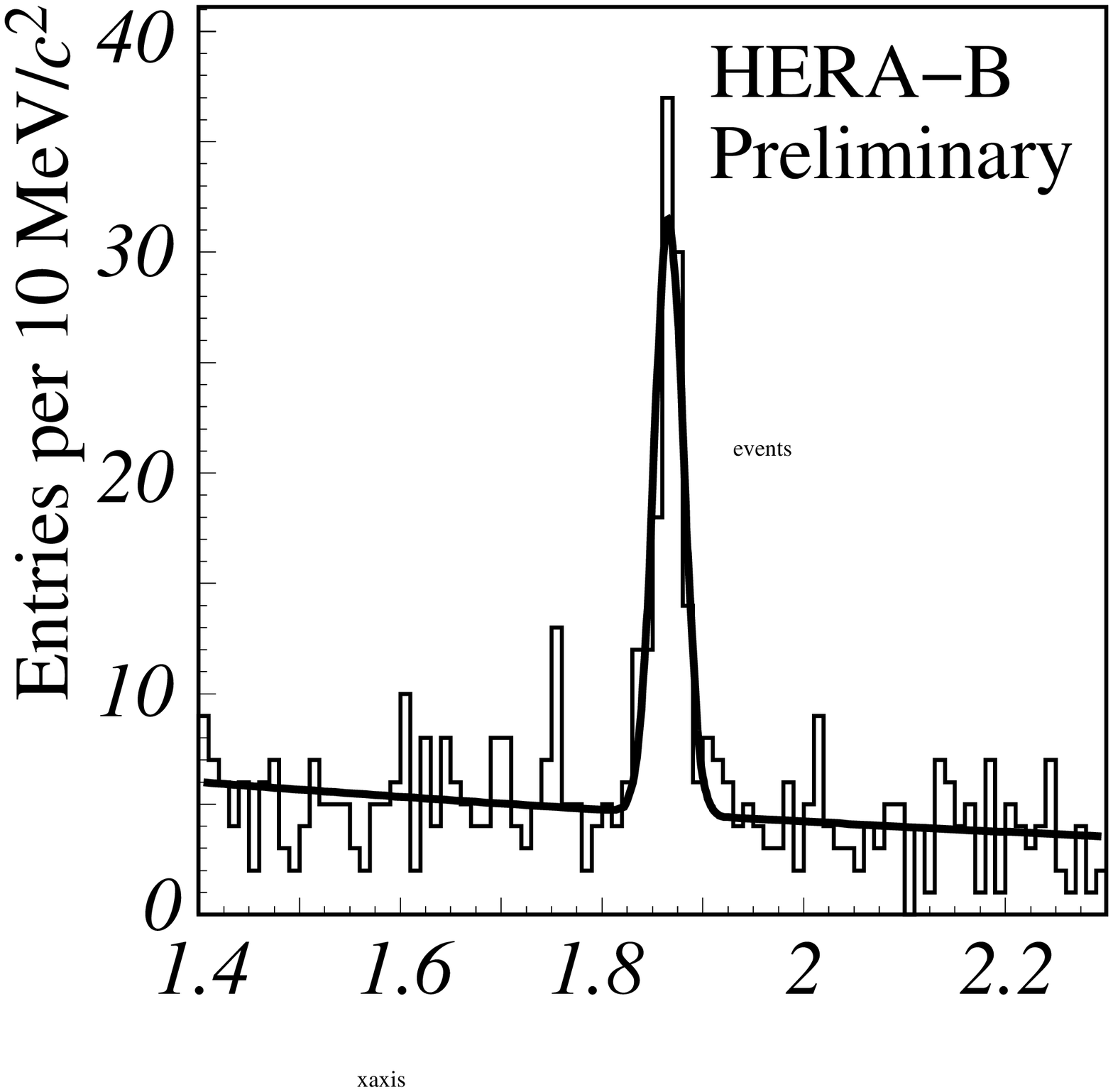}}
\hspace{3mm}
{\psfrag{events}{\small $43\pm8$ ev}
\psfrag{xaxis}{\small $M(\mathrm{K}\pi\pi)-M(\mathrm{K}\pi)$ [GeV/$c^2$]}\includegraphics[width=0.3\textwidth]{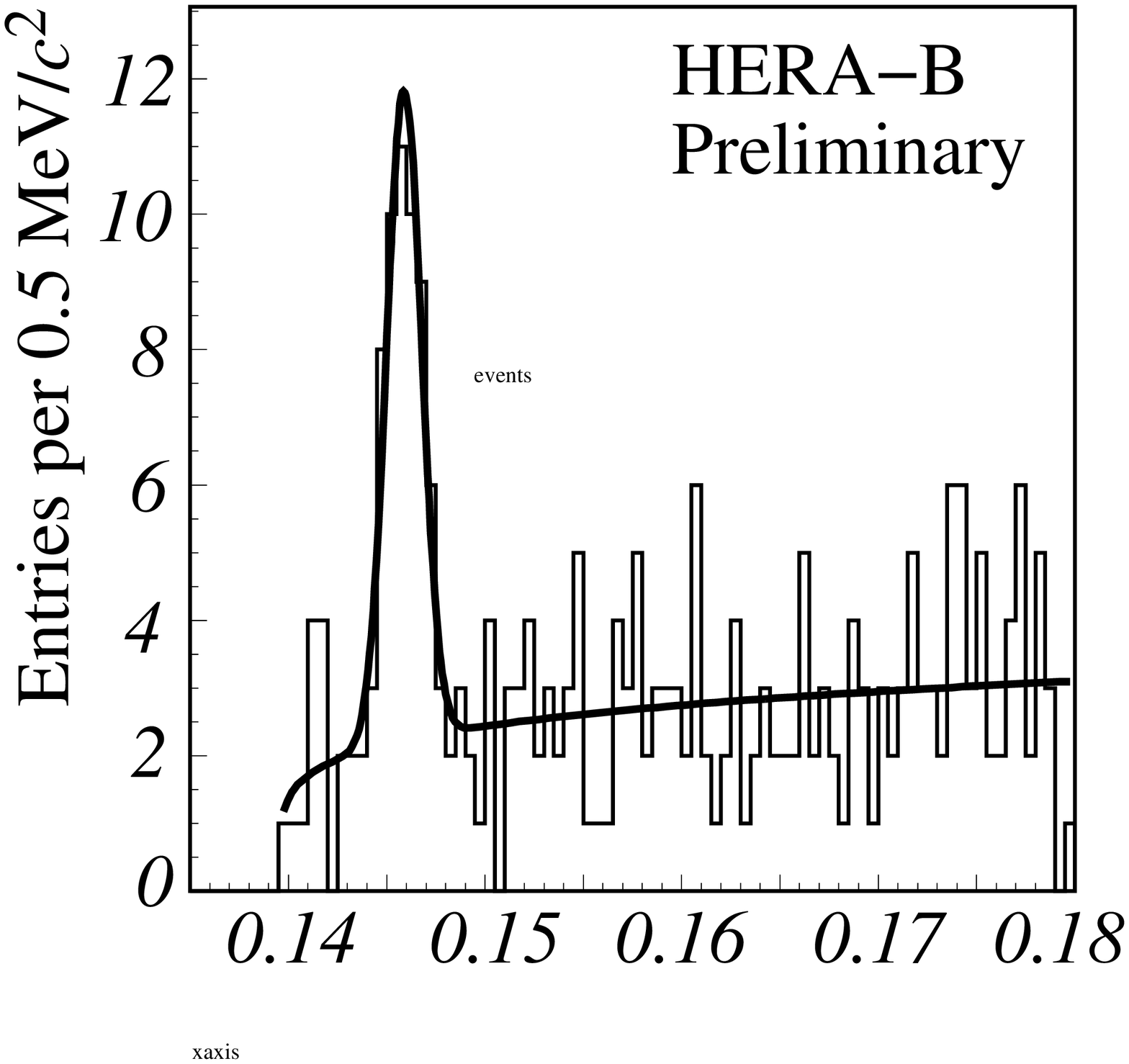}}
\vspace{-10mm}

\caption{Open charm decays observed in the minimum bias data-set: 
  $\PDz \rightarrow \PKm \pi^+$ (left), 
  $\PDp \rightarrow \PKp \pi^+ \pi^-$ (middle), 
  and $\PDstarp \rightarrow \PDz \pi^+ \rightarrow \PKm \pi^+ \pi^+$ (right).
The corresponding charge-conjugated channels are included.}
\label{fig:dsignal}
\end{figure*}

%
D meson production is studied using minimum bias triggered data.
The signals shown in Fig.~\ref{fig:dsignal} are obtained imposing cuts
on the separation of the decay vertex from the primary vertex and
the impact parameter to the target wire. 
 

\begin{figure}[t]
\centering
\includegraphics[width=0.33\textwidth]{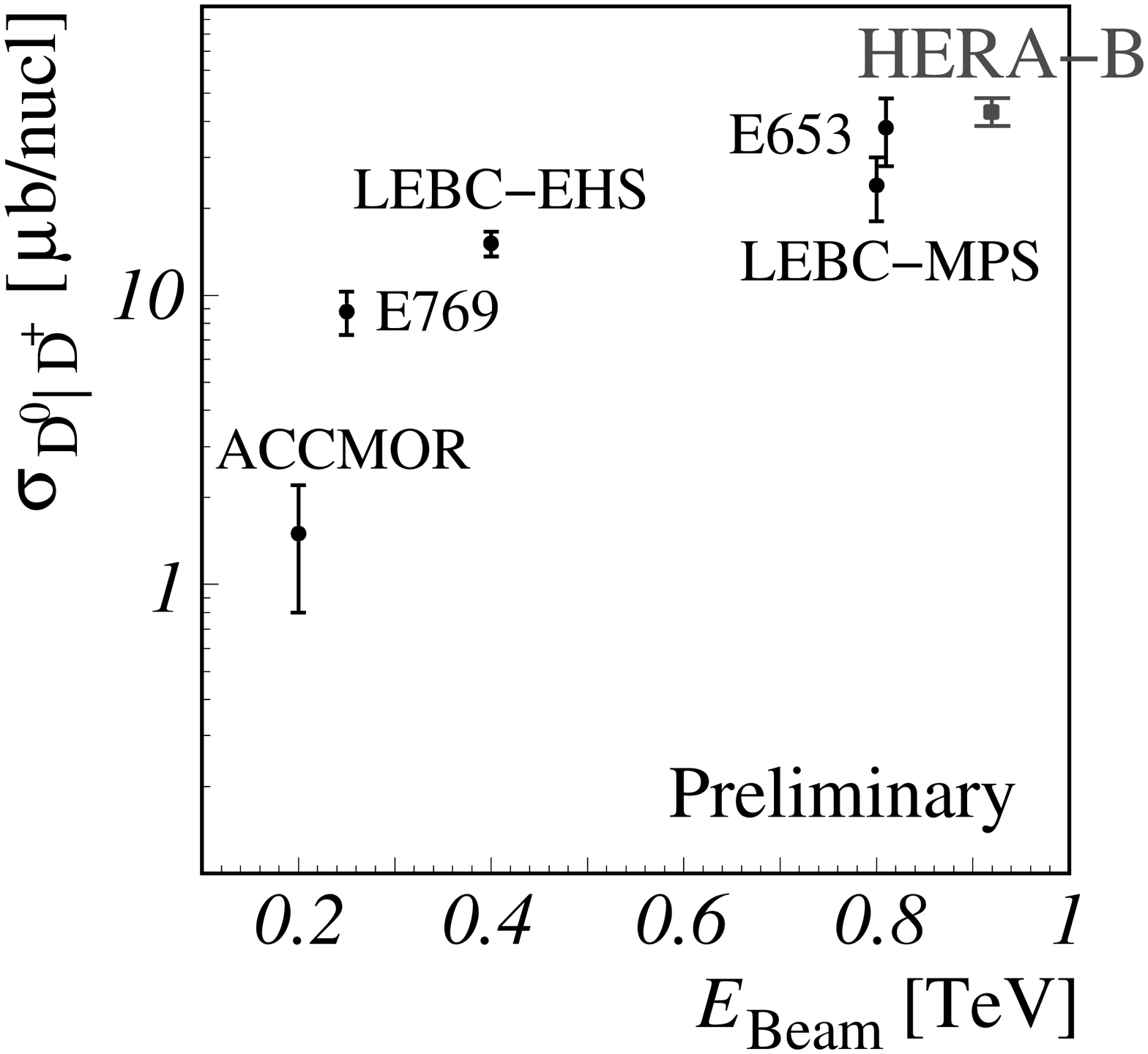}
\vspace{-1mm}

\includegraphics[width=0.33\textwidth]{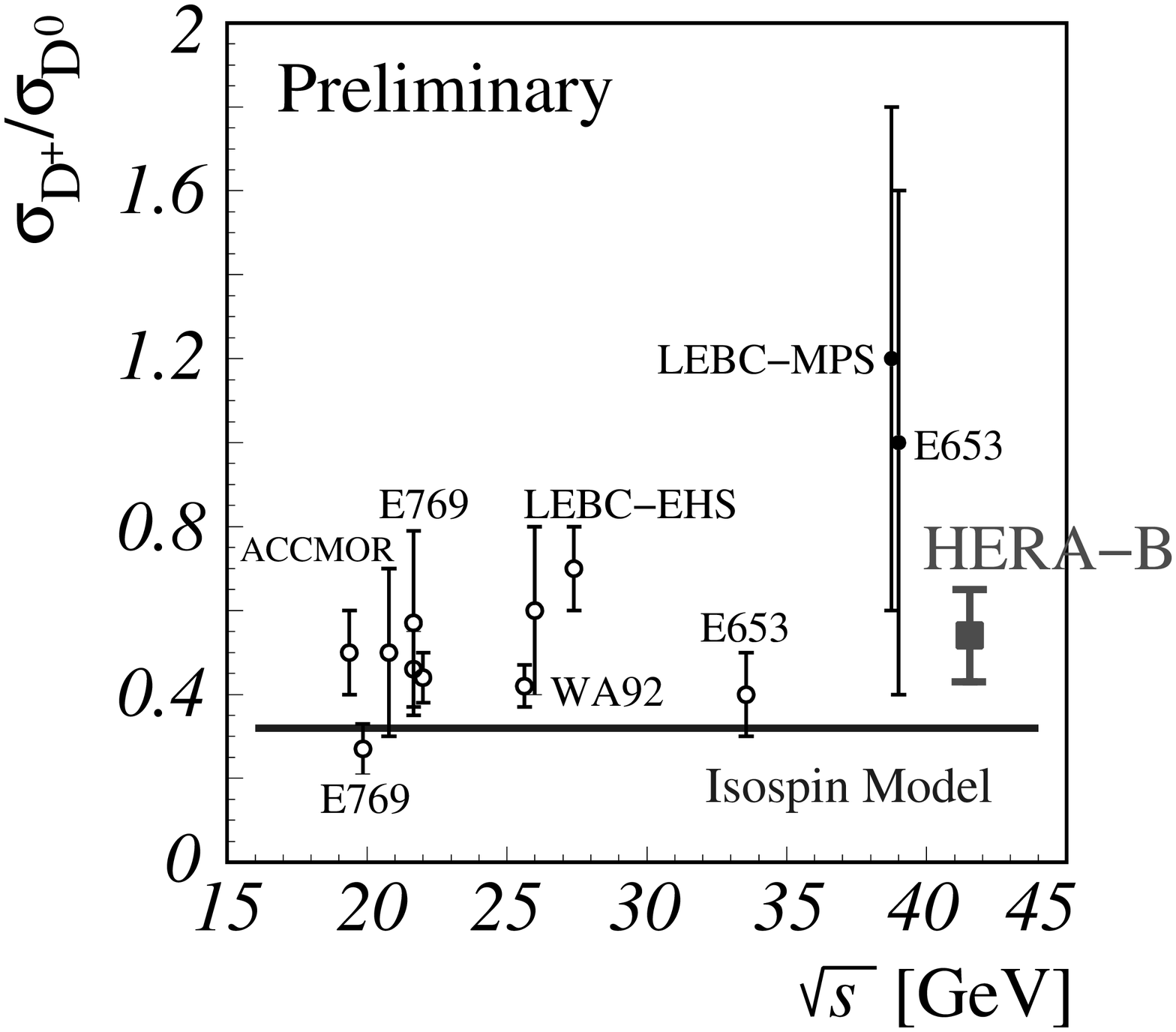}
\vspace{-10mm}

\caption{$\PDz$ and $\PDp$ production cross sections (top) and 
  cross section ratio (bottom), compared to results of previous
  experiments. 
Pion-induced results are depicted as open circles, while
  closed circles stand for proton-induced results.
  The line indicates a prediction for the
  production ratio based on an isospin model~\cite{Frixione:1998ma}.}
\label{fig:dcross}
\end{figure}

Using these data, the production cross section for $\PDz$ and $\PDp$,
$\sigma_{\PDz|\PDp} \equiv (\sigma_\PDz + \sigma_\PDp)/2$, is determined,
and the production ratios $\sigma_\PDp / \sigma_\PDz$ and  
$\sigma_\PDstarp / \sigma_\PDz$ are extracted. The cross sections obtained 
in the limited acceptance of the \hb{} detector are extrapolated to
the full \xf{} range using Pythia.  Preliminary results of these
measurements are summarized in Table~\ref{table:dcross}.
As shown in Fig.~\ref{fig:dcross}, the result on 
$\sigma_{\PDz|\PDp}$ agrees well with other
measurements at similar beam energies. The \hb{} measurement of the  
ratio  $\sigma_\PDp / \sigma_\PDz$  is in good agreement with pion-induced
results at lower center-of-mass energies. \hb{} will improve the results
of previous proton-induced experiments at similar energies.

\begin{table}
\caption{Production cross sections and cross section ratios for 
  \PDz, \PDp, and \PDstarp{} decays. The cross sections are given with
  statistical and systematic uncertainties.}
\label{table:dcross}
\small
\begin{tabular}{lcc}
\toprule
Preliminary & $-0.1 < \xf < 0.05$ & Full \xf{} range \\
\midrule
$\sigma_\PDz\unit{[\mu b/nucl]}$ & $21.4\pm3.2\pm3.6$ & $56.3\pm8.5\pm9.5$ \\
$\sigma_\PDp\unit{[\mu b/nucl]}$ & $11.5\pm1.7\pm2.2$ & $30.2\pm4.5\pm5.8$ \\
$\sigma_\PDstarp\unit{[\mu b/nucl]}$ & $10.0\pm1.9\pm1.4$ & $27.8\pm5.2\pm3.9$ \\
\midrule
$\sigma_\PDp/\sigma_\PDz$ & & $0.54\pm0.11\pm0.14$\\
$\sigma_\PDstarp/\sigma_\PDz$ & & $0.54\pm0.12\pm0.10$\\
\bottomrule
\end{tabular}
\end{table}

%
\subsection{Limit on $\PDz \rightarrow \dimuon$}
%
\begin{figure}[t]
\centering
\includegraphics[width=0.25\textwidth]{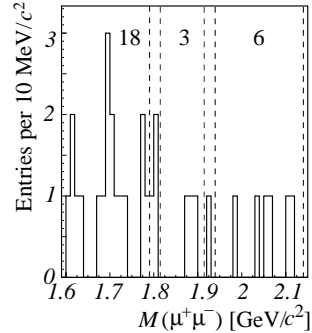}
\vspace{-8mm}

\caption{Dimuon invariant mass spectrum in the \PDz{} region. The numbers
of events in the signal region (middle)  and sideband regions 
are given~\cite{Abt:2004hn}.}
\label{fig:d0mumu}
\end{figure}

An upper limit on the branching fraction 
of the flavor-changing neutral current process 
$\PDz \rightarrow \dimuon$ has been determined
analyzing data collected with the dimuon trigger.
As shown in 
Fig.~\ref{fig:d0mumu}, three events are observed in the signal region,
while the expected background amounts to $6.0\pm1.2$ events. 
From these numbers, an upper limit for the $\PDz \rightarrow \dimuon$ 
branching fraction of $2.0\times 10^{-6}$ is obtained at 
90\% confidence level~\cite{Abt:2004hn}.

\section{Summary}
Measuring the properties of open and hidden charm production in proton-nucleus 
interactions is one of the main physics goals of the \hb{} experiment.
This paper presents preliminary results on a variety of physics studies,
including open charm and charmonium production cross sections
in proton-nucleus collusions,
differential distributions and nuclear effects. The \hb{} collaboration is
looking forward to finalizing these promising analysis efforts in the
near future.

\section{Acknowledgements}
The author would like to thank the organizers of BEACH04
for the fruitful and stimulating conference. 
This work was supported by the German BMBF under the contract number 5HB1PEA/7.
Special thanks to the
\hb{} charmonium and open charm working groups for 
helping me preparing this paper
and to DESY for the kind financial support.

\end{document}